\font\twelvebf=cmbx12
\font\ninerm=cmr9
\nopagenumbers
\magnification =\magstep 1
\overfullrule=0pt
\baselineskip=18pt
\line{\hfil }
\line{\hfil July 1998}
\line{\hfil PSU/TH/204}
\vskip .8in
\centerline{\twelvebf  On the Space-Time Uncertainty Principle and  Holography
}
\vskip .5in
\centerline{\ninerm D.MINIC}
\centerline{Physics Department}
\centerline{Pennsylvania State University}
\centerline{University Park, PA 16802}
\centerline {minic@phys.psu.edu}

\vskip 1in
\baselineskip=16pt
\centerline{\bf Abstract}
\vskip .1in

In this note further evidence is collected in support of
the claim that the
 space-time uncertainty principle implies
holography, both within the context of Matrix Theory and the
framework of the proposed duality between certain conformal field
theories and M-theory/string theory on AdS backgrounds.

\vfill\eject

\footline={\hss\tenrm\folio\hss}

\magnification =\magstep 1
\overfullrule=0pt
\baselineskip=22pt
\pageno=1

{\bf 1. Introduction}

The holographic principle [1] asserts that quantum theory of
gravity should be described via a boundary theory.
In particular, the number of dynamical degrees of freedom should satisfy the
Bekenstein-Hawking bound [2].

The background dependent proposals for a non-perturbative formulation
of M-theory, such as Matrix theory [3] (M-theory in the infinite
momentum frame), or proposals for a non-perturbative formulation
of IIB string theory on $AdS_5 \times S^5$ (and similarly
M-theory on $AdS_4 \times S^7$ and $AdS_7 \times S^4$) [3,4,5,6]
satisfy the Bekenstein-Hawking bound on the number
of degrees of freedom as required by the holographic principle [3, 6]. (The
boundary conformal field
theories (CFT) in the latter case live on the boundary of the
corresponding $AdS$ space).

It is our aim in this note to collect further evidence in support
of the claim, recently presented by Li and Yoneya [8],
which states that the holographic principle
is implied by the space-time uncertainty principle [7,8,9] both in
the context of Matrix theory and CFT/AdS duality. As it turns out,
 in both cases, the
space-time
uncertanty principle insures that the
ultraviolet behavior of the boundary Dp-brane [10] world-volume theory
is directly related to the infrared behavior of the theory in the
bulk [7,8,9], which in turn implies
holography.

{\bf 2. Space-time Uncertainty Principle in
String Theory and M-theory}

First we briefly review the space-time uncertainty principle
following the original work of Yoneya [7]. For a more complete
account the reader is referred to
 the recent beautiful survey of Li and Yoneya [8].

The origin of the space-time uncertainty
principle in string theory
 is in the time-energy uncertainty relation of quantum
mechanics [7,8]
$\Delta T \Delta E \geq \hbar $.
 In string theory, the energy is approximately proportional to
the string length $X$, so that
$\Delta E \sim {\hbar \over {\alpha'} } \Delta X
\sim{ \hbar \over {l_s}^2} \Delta X $. 
Thus the following space-time uncertainty relation is valid
$$
\Delta T \Delta X \geq {l_s}^2 . \eqno(1)
$$
The above formulation of the space-time uncertainty principle can
be related
to the opposite scaling of the longitudinal (including time) and
transverse directions of a string (or a
Dp-brane in general) [7, 8]. Short scales (UV) on the world-volume correspond
to long scales (IR) in target space. In other words, if the
space-time coordinates $X_i({\sigma}_m)$ scale as
$$
X_i(\sigma_m) \rightarrow \lambda X_i(\sigma_m) , \eqno(2)
$$
then the world-volume coordinates $\sigma_m$ must scale inversely as
$$
\sigma_m \rightarrow \lambda^{-1} \sigma_m . \eqno(3)
$$

In [7,8] it has been also pointed out
that the space-time uncertainty
principle in string theory is intimately related
to the original Regge/resonance duality (s-t channel
duality of the old dual resonance models).
For example, the Regge regime corresponds to the limit
$\Delta X \rightarrow \infty$. On the other hand, the
resonance regime corresponds to the limit  $\Delta T \rightarrow \infty$.

The space-time uncertainty principle of
the theory of closed strings is also related to the existence of
a massless spin 2 particle in the spectrum [7,8]. The argument
goes as follows:
high-energy scattering gedanken experiments probe
 the short time scales; hence,
$\Delta T \rightarrow 0$. The amplitude of the scattering
process is proportional
to $\Delta X_{long} \sim l_{s}^2 /{\Delta T} \sim E $,
which in turn implies the existence of a massless spin 2 particle, the
graviton.
(The intercept of the leading Regge trajectory $\alpha(t)$
is 2, which follows from the dependence of the scattering
amplitude on energy, $E^{\alpha(t) -1} = E$.)

On the other hand, the region $\Delta T \rightarrow \infty$, which according to
the space-time uncertainty principle corresponds
to $\Delta X \rightarrow 0$, is associated with
 short distance processes
in  string theory, i.e. with  the physics of D-branes [9,10,11].
It has been also pointed out by Yoneya 
that the space-time uncertainty principle is
closely related
to the fact that perturbative string theory is conformally invariant [7].

Furthermore, Li and Yoneya [9] showed that the description of Dp-branes in
terms of $U(N)$ super Yang-Mills theories is aslo compatible
with the space-time uncertainty principle.
In this case one examines the behavior of the low energy
theory of $N$ D0-branes, i.e. the action of the ${\cal{N}}=16$ $U(N)$
 supersymmetric
Yang-Mills quantum mechanics [3, 12] (with $l_s =1$)
$$
S = \int dt Tr {1 \over g_s} ({1 \over 2}{({D_t{X}}_{a})}^2
+ i \bar{\theta} D_t \theta +{1 \over 4} [X_a,X_b]^2
- \bar{\theta} \Gamma^{a} [\theta,X_a] ) . \eqno(4)
$$
Here
$D_t X_a = \partial_t +[A_0,X_a]$.
As Li and Yoneya [9] noticed, this expression is invariant under (in 
 $A_0 =0$ gauge)
$$
X \rightarrow g_{s}^{1/3} X, \quad t \rightarrow g_{s}^{-1/3} t ,
\eqno(5)
$$
which is consistent with (1).
The scale $g_{s}^{-1/3} l_s$ is nothing but the 11-d Planck length of
Matrix theory [3].
There are many nice physical applications of
(5) [9]. For example, the order of magnitude estimate of the
distances probed by D0-branes is given by $\Delta X \sim \sqrt{v} l_s$
(this follows from $\Delta T \Delta X \sim
{ {(\Delta X)^{2}} \over v} \sim \l_s^2$). The spreading of a
D0-brane wave packet is estimated from $\Delta X \sim \sqrt{v} l_s$
and $X \sim g_{s}^{1/3} l_s$ to be of the order of $g_s l_s/v$.
All these results can be also obtained from the Born-Oppenheimer
approximation for the coupling between the diagonal and
off-diagonal $U(N)$ matrix elements as in [11].

The argument of Li and Yoneya [9] can be generalized to other Dp-branes.
The longitudinal and transverse lengths scale oppositely with the
powers of the string coupling constant ${g_s}^{-1/(3-p)}$ and
${g_s}^{1/(3-p)}$ (for $p=3$ the
super Yang-Mills theory is conformally invariant).
 As pointed out by Yoneya [7], these scalings are
consequences of the fact that the interactions between Dp-branes are
mediated via open strings, whose respective longitudinal and
transverse lengths satisfy the space-time uncertartainty relation.

The mathematical structure behind the space-time uncertainty principle is
at the moment unknown.
The simplect Lorentz covariant formulation
of the space-time uncertainty principle, as proposed by
Yoneya [7], is to assume non-commutativity of
all space-time coordinates
$$
[X_{\mu}, X_{\nu}]^2 \sim I .  \eqno(6)
$$
This relation leads immediately to eq. (1).
Unfortunately, at the moment, there is no
satisfactory covariant formulation of M-theory which would
incorporate this statement of the space-time uncertainty
principle. (See [13] for some preliminary comments regarding this
question.)

But the space-time uncertainty principle in string theory can be
generalized to M-theory.
As shown in [8], the
 space-time uncertainty relation of Matrix theory reads 
$$
\Delta X_{-} \Delta X_{a} \Delta X_{+} \sim l_{11}^{3} . \eqno(7)
$$
Here
$X_{-}$ denotes the longitudinal direction and $X_{+}$ - the
global time of Matrix theory.
This follows from the fact that $\alpha' = l_{11}^{3}/R$, $l_{11}$ being the
11-d Planck length. In Matrix theory
$\Delta X_{-} \sim R$, the extent of the eleventh dimension; hence,
(7) is compatible with (1).
Ideally, the M-theory space-time uncertainty relation (7)
 should be used as a guide towards
a covariant formulation of Matrix theory.

Here we wish to point out that (7) is consistent with the covariant formulation
of 11-d membrane [12],
if the latter is suitably discretized (for example, along the
lines of [13]) in order to take into
account the fact that the membrane world-volume should be at least of
order $l_{11}^{3}$.

First (7) implies the following scaling of space-time coordinates
$$
X_{\mu} \rightarrow g^{1/3} X_{\mu} . \eqno(7a)
$$
Notice that this relation is consistent 
with the classical membrane Nambu-Goto action
$S_{mem} \sim {1 \over g}\int d^{3}\xi \sqrt{-detg_{ij}} $, 
where $g_{ij} = \partial_{i} X^{\mu}
\partial_{j} X^{\nu} \eta_{\mu \nu}$ is the induced metric on
the 3-volume of the membrane
(for simplicity we consider only the bosonic case as in [13]).
In light cone-gauge [12], (7a) reduces to (5).

Second, note that the M-theory space-time uncertainty relation  (7) is
consistent with the energy-time uncertainty relation, once
we take into account that the energy of the membrane is proportional to its
area.  Of course, strictly speaking this argument is not correct, because
the membrane is an effective object in M-theory. On the other
hand this should be true for the
case of a discretized covariant membrane [13], 
whose volume is bounded from below
by $l_{11}^{3}$.

Finally, note that (7) implies a relativistic bound on the uncertainty
of the transverse coordinate
$\Delta X \sim \sqrt{c/{\Delta X_{-}}} \quad l_{11}^{3/2}$, 
(c is the velocity of light),
once we take into account
that $c \Delta X_{+} \sim \Delta X$. 
One expects, in analogy with quantum field theory where
the uncertainty for the coordinate of a relativistic particle 
corresponds to a momentum uncertainty proportional to the
threshold energy for particle-antiparticle production, that
 processes which lead to a bound on $\Delta X$ should
 involve brane-anti-brane interactions, and should be described in
any covariant version of Matrix theory consistent with (7).

{\bf 3. Space-time Uncertanty Principle, Holography and Matrix Theory}

 The point of this section
is to discuss evidence in favor of
the claim that the space-time uncertainty principle
implies holographic behavior
in Matrix theory [8].

We start by reviewing the work of Jevicki and Yoneya [14].
First, these authors show that there is a hidden $SU(1,1) $
symmetry of the Matrix theory action (4). The symmetry is
generated by scale transformations (D) (implied by
the space-time uncertainty relation), time translations (H),
 and
special conformal transformations (K) (under which the string constant,
treated as a part of background field variables,
also changes [14]).

The explicit actions of D, H and K are given by [14]
$$\eqalign{
\delta_D X_a = X_a, \quad \delta_D A_0= A_0, \quad \delta_D t = -t,
\quad \delta_D g_s = 3g_s  \cr
\delta_H X_a = 0, \quad \delta_D A_0= 0, \quad \delta_H t = 1,
\quad \delta_H g_s = 0 \cr
\delta_K X_a = 2t X_a, \quad \delta_K A_0= 2t A_0, \quad \delta_K t = -t^2,
\quad \delta_K g_s = 6t g_s .
}
\eqno(8a)
$$
These transformations form an $SU(1,1)$ algebra
$$
[\delta_D, \delta_H] = \delta_H, \quad
[\delta_D, \delta_K] = -\delta_K, \quad
[\delta_H, \delta_K] = 2\delta_D. \quad
\eqno(8b)
$$
Furthermore, Jevicki and Yoneya [14] demonstrate
that the above conformal symmetry
(which stems from the space-time uncertainty principle) taken together with
the SUSY non-renormalization theorem [15] implies the lagrangian
of Becker-Becker-Polchinski-Tseytlin (BBPT) [16], which is an
effective lagrangian of a D0-brane probe in the background
of heavy D0-brane sources. (The BBPT lagrangian [16] is
computed in the framework of a discrete light-cone
formulation of Matrix theory [17], where the number of
partons $N$ is kept finite.) This lagrangian is valid in the limit of
large distances and small relative velocities [16] (with $l_s=1$)
$$
S_{eff} = - \int dt {1 \over {g_s h(r)}}  (\sqrt{1-h(r) v^2} -1) ,
\eqno(9a)
$$
where the static D0-brane background is described by
$h(r) \equiv h_{--}(r) = {15 N \over {2g_{s}^5 r^{7}}}$.

Now, in the limit of mean field theory, the BBPT action implies 
holography. This particular fact
has been already demonstrated in the study of neutral black hole - like
bound states of D0-branes in the infinite momentum
frame [18]. The characteristic size of such a
bound state is essentially the Schwarzschild radius $R_h$. The
characteristic velocity of D0-branes in the infinite momentum
frame is determined by the inverse of the boosting parameter,
or basically from the Heisenberg uncertainty relation $v R_h \sim 1$.
The Bekenstein-Hawking scaling relation for the black-hole entropy 
 $S \sim R_h^{D-2}/G_{D}$,
in D-spacetime dimensions, follows after
the virial theorem is applied to (9a) (at the
transition point $S \sim N$ [18,19]).
Thus we conclude, though indirectly,
 that the space-time uncertainty principle together
with the SUSY non-renormalization theorem implies holography.

Note that
this picture can be, at least formally,
 generalized to the charged black holes discussed in [20]
(I thank S. Chaudhuri and M. Li for many discussions regarding this issue):
For a large boost the energy $E$ and momentum $P$ of a
charged black hole are according to [20]
$E \sim \mu A \exp{\alpha},  P \sim \mu A \exp{\alpha}$, following
the notation of [20]. Also, the black hole mass scales as
$M \sim \mu A \sim R_{h}^{D-3} C^{-1/2}$ where
$A(\beta_i) \equiv
(\lambda + \sum_{i=1}^n \cosh 2\beta_i)$ and
$C(\beta_{i}) \sim A(\beta_i)^{-2}  B(\beta_i)^{2{(D-3)}/{D-2}}$
and  $B(\beta_i) \equiv \prod_{i=1}^n \cosh\beta_i$, again 
following the notation of [20].
Thus, the black hole longitudinal momentum is determined as
$P \sim R_{h}^{D-3} C^{-1/2} \exp{\alpha}$.
Let us concentrate on the special point
$P = N/R$.
The boosting parameter $\exp{\alpha}$
 is given by
$\exp{\alpha} \sim C^{1/2} R_{h}/R $.
This parameter determines by how much the box of size $R$ has
to expand to accomodate a charged black hole [20] 
of the horizon radius $R_{h}$.
Suppose we postulate that the radius of the
bound state $r_b$ is related to the horizon radius $R_h$ via
$R_h \sim r_b C^{1 \over {D-4}}$. 
The physical meaning of this relation is simple: the line elements of [20]
roughly have the Reissner-Nordstr\"{o}m form, which implies that there
should exist two
characteristic scales describing the physics of charged black
holes.
The presence of charges is hidden in the complicated function
$C$, which we take as a phenomenological input.
The two scales naturally coalesce into one, $R_h$, if we consider
a neutral black hole.
The characteristic partonic velocity is determined by the inverse of
the boosting parameter $\exp{\alpha}$, that is,
$v \sim {R \over R_{h}} C^{-1/2}$.
Then the application of the virial theorem to the BBPT effective lagrangian
gives the Bekenstein-Hawking formula at the special point $N \sim S$,
i.e. $ S \sim N \sim {R_h^{D-2} \over G_D}$.
The virial theorem and the first law of thermodynamics imply
the equation of state (after we utilize the  Bekenstein-Hawking
scaling)
$S \sim (NT/R)^{{D-2}\over {D-4}} C^{{D-2}\over {D-4}}$.
This relation tells us that $R_h \sim (NTG_D/R)^{1 \over {D-4}}
C^{ 1 \over {D-4}}$, which is exactly the relation between the size of
the bound state $r_b$ and the horizon radius $R_h$, if the size of the
bound state is defined to scale as $r_b \sim (NTG_D/R)^{1 \over {D-4}}$.
Also, the infinite momentum dispersion
relation $E_{lc} \sim M^{2}R/N $ leads to
$\mu \sim (NT/R)^{(D-3)/(D-4)} (C^{{D-2} \over {2(D-4)}}/A) $, 
 which agrees with [20].
When $C \rightarrow 1$ we recover the familiar expression for the
Schwzrschild black hole, as expected.

Another way to see holography of (9a) is to realize
that the
BBPT effective action is nothing but the Einstein-Hilbert action
$S_{EH} = - {1 \over {16 \pi G}} \int dx^{11} \sqrt{g} R $, 
written in the infinite momentum frame [21], in the limit of
large distances and small relative velocities.
Then by formally applying the Gibbons-Hawking argument [22, 23] to the
euclidean
partition function defined by the above action
one finds that the entropy
satisfies the Bekenstein-Hawking scaling.
One might expect that this formal counting 
is valid in this case because one is only looking at the
low-energy behavior of Matrix theory.
But a word of caution is needed here.
Strictly speaking, we cannot use the 
Gibbons-Hawking argument given the current
dictionary between Matrix theory and 11-d supergravity [24].
In the linearized 11-d supergravity one cannot talk about
black hole bound states,
so the above scaling arguments do not apply.
On the other hand there exists a very beautiful argument due to
Jacobson [25], which essentially states
 that a holographic theory of gravity which satifies the
usual axioms of quantum field theory in curved space time, 
implies
the full non-linear structure of general relativity. In our case this
would translate into a statement that holography in Matrix theory (as implied
by the space-time uncertainty principle)
together with locality should 
lead to
 the full non-linear 
structure of 11-d supergravity.
But to be able to directly apply Jacobson's argument [25] we also need
a fully covariant version of Matrix theory.

Fortunately, we can do better than to observe holography in
Matrix theory in the background of
static D0-brane sources. Using the recent powerful results of
Okawa and Yoneya [26] on the full structure of 3-body interactions in
Matrix theory at finite $N$, we can argue that holography is valid even 
when many-body interactions of D0-branes are taken into
account.

Indeed, the effective lagrangian that describes 3-body interactions of
D0-branes according to Okawa and Yoneya has two pieces (eqs. 2.53 and 2.55 of
[26]):
$$
L_{V} = - \sum _{a,b,c} {{(15)^{2} N_{a}N_{b}N_{c}} \over {64 R^{5} M^{18}}}
v_{ab}^{2} v_{ca}^{2} (v_{ab} \cdot v_{ca}) r_{ab}^{-7} r_{ca}^{-7}
$$
and
$$\eqalign{L_{Y} =  &- \sum_{a,b,c}
{{(15)^{3}
 N_{a}N_{b}N_{c}} \over {96 (2 \pi)^{4} R^{5} M^{18}}}
[-v_{bc}^{2} v_{ca}^{2} (v_{cb} \cdot \nabla_{c}) (v_{ca} \cdot \nabla_{c}) \cr
&+ {1 \over 2} v_{ca}^{4} (v_{cb} \cdot \nabla_{c})^{2}
+ {1 \over 2} v_{bc}^{4} (v_{ca} \cdot \nabla_{c})^{2}
- {1 \over 2}
v_{ba}^{2} v_{ac}^{2} (v_{cb} \cdot \nabla_{c}) (v_{bc} \cdot \nabla_{b}) \cr
&+ {1 \over 4} v_{bc}^{4} (v_{ba} \cdot \nabla_{b}) (v_{ca} \cdot \nabla_{c})]
\Delta (a,b,c) \cr}$$
where
$\Delta (a,b,c) \equiv \int d^{9}y |x_{a}-y|^{-7}|x_{b}-y|^{-7}|x_{c}-y|^{-7}$
and $a,b,c$ are particle labels.

It is now easy to see that both $L_{V}$ and $L_{Y}$
lead to
 the Bekenstein-Hawking
scaling if the relative velocities of particles saturate the Heisenberg
uncertainty bound $v r \sim 1$ at the transition point $S \sim N$
 in the large N limit. The $L_{Y}$ term is much more subtle than $L_{V}$
 in this respect,
 but the scaling argument still works.
The
effective lagrangians
for 2- and 3-body interactions follow from the general structure of (4),
which is compatible with the space-time uncertainty principle, thus 
supporting the claim that the space-time uncertainty principle 
underlies holography in Matrix theory.

We end this section with a simple observation about the hidden
$SU(1,1) \sim SO(2,1)$ symmetry
in Matrix theory [14]: In [27] it was argued that in the
large N limit there exists an $SO(1,2) \times SO(16)$ symmetry
in Matrix theory. This should be compared to the well known fact that the
11-d supergravity can be formulated in such a way so that the same symmetry
is made manifest [28].
Notice that the 
$SO(16)$ symmetry is not 
an
isometry of
any compactification of the 11-d supergravity. Rather this
 $SO(16)$ naturally
appears as
a local symmetry of the 11-d supergravity dimensionally reduced to
3-d [29].

{\bf 4. Space-time Uncertanty Principle, Holography and AdS/CFT Duality}

In this section we
comment on the claim
 that the space-time uncertainty principle implies holography [8] in
the context of CFT/AdS duality [30].

The argument of Susskind and Witten [30] relies on the connection
between the UV behavior of the boundary conformal field
theory and the IR behavior of the bulk AdS theory
of gravity.
In fact, this UV-IR relation of Susskind and Witten
is nothing but the statement of the space-time uncertainty
principle, as we have seen in section 1. (eqs. (2) and (3)).

The argument for the holographic bound goes as follows [30]:
The number of degrees of freedom $N_{dof}$ per unit volume in the boundary
CFT theory (${\cal{N}}=4$ super $SU(N)$ Yang-Mills in 4-d) scales as
$$
N_{dof} \sim N^2/{\delta}^3, \eqno(10)
$$
where $\delta$ is an
UV cut-off of the world-volume $CFT_4$ theory which
lives on the boundary of $AdS_5$.
The spatial volume of the boundary theory is $V_3=R_{b}^3/{\delta}^3$
which implies
$$
N_{dof} \sim V_3 N^2/R_{b}^3. \eqno(11)
$$
But CFT/AdS duality, following
the original work of Maldacena [4],
relates
$N$ and the radius of $AdS_5$
$$
R_{b} = l_s (g_s N)^{1/4} ,  \eqno(12)
$$
which in combination with eq. (10) gives
the Bekenstein-Hawking bound.

The argument of [30] can be readily generalized for the case of
membranes and five branes [8]. The relevant number of degrees of
freedom in the former case should scale as $N^{3/2}$ and in the latter
as $N^{3}$. The latter scaling might
be naturally interpreted if we had
a non-abelian formulation of theories involving two-forms, which we don't.

But the
 relationship between $N^{3/2}$ and $N^{3}$ could be understood
from the following picture:
Imagine a covariant formulation of Matrix theory in which the world-volume
of the M-theory membrane is suitably
discretized, as described in section 2. The
covariant formulation should also naturally include the covariant
M-theory five-brane.
Given the fact that in Matrix theory [31] the longitudinal five-brane wrapped
around the longitudinal direction can be seen by stacking
two orthogonal
transverse membranes, we can envision stacking two orthogonal
discretized
membrane
world-volumes in covariant Matrix theory to get a covariant
M-theory five-brane.
Then in order to match the energy densities of two objects
the number of degrees of freedom of the theory that describes the discretized
membrane should be square root of the number of degrees of freedom that
describe the covariant M-theory five-brane. Hence $N^{3}$ degrees of
freedom of the five-brane theory imply $N^{3/2}$ degrees of freedom of
the membrane theory, as it should be.

Actually, we can extend the argument Susskind and Witten, to include
conformal field theories
in other number of dimensions. (This has been
noticed in conversations with M. Li.)
For the cases considered above the number of degrees of freedom in d-dimensions
goes as $N^{d/2}$. Then,
  there should
exist a superconformal quantum mechanics with $N^{1/2}$ degrees of freedom,
and a five dimensional superconformal theory with $N^{5/2}$ degrees of
freedom. The two dimensional superconformal field theory with $N$
abelian degrees of freedom also fits this pattern. The argument of
Susskind and Witten which gives Bekenstein-Hawking scaling in the end,
works nicely in each of these cases.

{\bf 5. Background Independent Holography via 2-Hilbert spaces?}

The above discussion of the space-time uncertainty relations and
holography is obviously background dependent. (One possible exception
to this statement
is the theory proposed by Ho\v{r}ava [32].)
In conclusion to this note, we wish to point out that  background independent
holography might
be understood via the concept of 2-Hilbert spaces, in analogy with
a kinematical set-up proposed by Crane [33] in relation
to $3+1$ dimensional quantum general
relativity (see the article by Smolin [33]
for a nice review).

The main idea of Crane [33] is
that a background independent holographic theory,
such as quantum gravity, should not be described in terms of
a single Hilbert space, but in terms of a linear structure which is
spanned by basis elements which are also taken to be
Hilbert spaces (such structures are
called 2-Hilbert spaces; see the review of Baez [33]
for the
precise mathematical set-up within the
framework of category theory.) 

Very roughly, according to Baez [33],
 the basic objects of
2-Hilbert spaces are finite dimensional Hilbert spaces (replacing
vectors as the basic elements of Hilbert spaces). The zero object (the analog
of the zero vector) is a zero-dimensional Hilbert space; the analog of
adding two vectors is forming the direct sum; the analog of multiplying
a vector by a complex number is  tensoring an object by a Hilbert space, and
the analog of the inner product of two vectors is a bifunctor taking
each pair of objects $a$, $b$ of a 2-Hilbert space to the set of morphisms
from $a$ to $b$.

Thus, following Crane,
 the background independent 2-Hilbert space of the bulk quantum theory
of gravity should be constructed in terms of component Hilbert spaces of the
appropriate
boundary theories.
We want to illustrate this idea within the framework of Matrix theory:
A Hilbert space in Matrix theory is represented by block diagonal
matrices [3]. Denote such a Hilbert space by $H_{bound}^{i}$.
In order to realize the idea of Crane [33] we
suppose that each of these Hilbert spaces can be taken as
a component of a 2-Hilbert space ($Hilb$). In other words, an element $H_2$
of $Hilb$
can be written as a linear combination of $H_{bound}^{i}$
$$
H_2 = \sum_{i} c_{i} H_{bound}^{i}.  \eqno(13)
$$
Now,  
let each $H_{bound}$ satisfy the
 holographic bound (i.e. let the dimension of a particular
 $H_{bound}$ be determined
 by the appropriate boundary theory).
This implies the holographic bound 
on the dimension of the
bulk 2-Hilbert space,
and therefore - background independent holography.
Coming back to our membrane analogy from section 2.,
the corresponding 2-Hilbert
space can be concretely
envisioned as a collection of transverse membranes, each
 representing the Hilbert
space of a particular background.

{\bf Acknowledgements:} I wish to thank 
I. Bars, S. Chaudhuri, M. G\"{u}naydin,
 M. Li, T. Yoneya, J. Polchinski and L. Smolin for
useful discussions. I am especially indebted to M. Li and T. Yoneya
for sharing their insights on the space-time uncertainty principle and
holography with me. Extra thanks to M. Li for providing a reliable source
of friendly criticism and for commenting on the preliminary version of this
note.

\vskip.1in
{\bf References}

\item{1.} G. 't Hooft, gr-qc/9310026; L. Susskind, Jour. Math. Phys. 36
(1996) 6377.
\item{2.} J. D. Bekenstein, Phys. Rev. D7 (1973) 2333;
S. W. Hawking, Comm. Math. Phys, 43 (1975) 199;
\item{3.} T. Banks, W. Fischler, S. H. Shenker and L. Susskind, Phys. Rev. D55
(1997) 5112. For a review and further references, see T.Banks, hep-th/9706168;
D. Biggati and L. Susskind, hep-th/9712072; W. Taylor, hep-th/9801182.
\item{4.} J. Maldacena, hep-th/9711200.
\item{5.} S. S. Gubser, I. Klebanov and A. M. Polyakov, hep-th/9802109;
\item{6.} E. Witten, hep-th/9802150; hep-th/9803131.
\item{7.} T. Yoneya, Mod. Phys. Lett. A4 (1989) 1587; See also
T. Yoneya, in "Wandering in the Fields". K. Kawarabajashi and A. Ukawa, eds.
(World Scientific, 1987) pp.419; and "Quantum String Theory", N. Kawamoto and
T. Kugo, eds. (Springer, 1988) pp.23.
T. Yoneya, Prog. Theor. Phys. 97 (1997) 949; hep-th/9707002
\item{8.} M. Li and T. Yoneya, hep-th/9806240.
\item{9.} M. Li and T. Yoneya, Phys. Rev. Lett. 78 (1997) 1219.
\item{10.} J. Polchinski, Phys. Rev. Lett. 75 (1995) 4724;
J. Dai, R. G. Leigh and J. Polchinski, Mod. Phys. Lett. A4 (1989) 2073;
J. Polchinski, {\it TASI Lectures on D-branes}, hep-th/9611050.
\item{11.} M. R. Douglas, D. Kabat, P. Pouliot and
S. H. Shenker, Nucl. Phys. B485 (1997) 1219.
\item{12.} For references consult
B. de Wit, J. Hoppe and H. Nicolai, Nucl. Phys. B305 (1988) 545.
\item{13.} H. Awata and D. Minic, JHEP 04 (1998) 006.
\item{14.} A. Jevicki and T. Yoneya, hep-th/9805069.
\item{15.} M. Dine, R. Echols and J. Gray, hep-th/9805007;
   S. Paban, S. Sethi and M. Stern, hep-th/9805018;hep-th/9806028 .
\item{16.} K. Becker and M. Becker, hep-th/9705091;
K. Becker, M. Becker, J. Polchinski and
A. Tseytlin, hep-th/9706072.
\item{17.} L. Susskind, hep-th/9704080;
N. Seiberg, Phys. Rev. Lett. 79 (1997)
3577; A. Sen, hep-th/9709220; S. Hellerman and J. Polchinski, hep-th/9711037.
\item{18.} T. Banks, W. Fischler, I. Klebanov and L. Susskind,
hep-th/9709091;  hep-th/9711005; I. Klebanov and L. Susskind, hep-th/9709108;
E. Halyo, hep-th/9709225; G. Horowitz and E. Martinec, hep-th/9710217;
 M. Li, hep-th/9710226;
 T. Banks, W. Fischler and I. Klebanov, hep-th/9712236;
 H. Liu and A. Tseytlin, hep-th/9712063; D. Minic, hep-th/9712202;
 N. Ohta and J.-G. Zhou, hep-th/9801023; M. Li and E. Martinec, hep-th/9801070;
 D. Kabat and G. Lyfshytz, hep-th/9806214.
\item{19.} S. Chaudhuri and D. Minic, hep-th/9803120;
B. Sathialpan, hep-th/9805126; L. Susskind, hep-th/9805115.
\item{20.} M. Cveti\v{c} and A. Tseytlin, Nucl. Phys. B478 (1996) 181.
\item{21.} E. Keski-Vakkuri and P. Kraus, hep-th/9709122;
hep-th/9711013.
\item{22.} D. A. Lowe, hep-th/9802173.
\item{23.} G. W. Gibbons and S. W. Hawking, Phys. Rev. D15 (1977) 2752.
\item{24.} D. Kabat and W. Taylor, hep-th/9712185;
I. Chepelev and A. Tseytlin, hep-th/9709087; hep-th/9704127.
\item{25.} T. Jacobson, Phys. Rev. Lett. 75 91995) 1260.
\item{26.} Y. Okawa and T. Yoneya, hep-th/9806108.
\item{27.} M. G\"{u}naydin and D. Minic, hep-th/9702047.
\item{28.} H. Nicolai, Phys. Lett. B187  (1987) 316; hep-th/9801090.
\item{29.} N. Marcus and J. H. Schwarz, Nucl. Phys. B 228 (1983) 145.
\item{30.} L. Susskind and E. Witten, hep-th/9805114.
\item{31.} T. Banks, N. Seiberg and S. Shenker, hep-th/9612157.
\item{32.} P. Ho\v{r}ava, hep-th/9712130.
\item{33.} L. Smolin, Jour. Math. Phys. 36 (1996) 6417, gr-qc/9505028;
L. Crane, Jour. Math. Phys 36 (1996) 6180;
D. Freed, Comm. Math. Phys. 159 (1994) 343;
J. Baez, Adv. of Math, 127 (1997) 125.

\end